\documentclass[journal=jacsat,manuscript=article]{achemso}
\usepackage{chemformula} 
\usepackage[T1]{fontenc} 
\usepackage{atbegshi}
\AtBeginDocument{\AtBeginShipoutNext{\AtBeginShipoutDiscard}\addtocounter{page}{-1}}

\newcommand{\BiSe}{Bi$_2$Se$_3$}

\newcommand{\jh}[1]{\textcolor{black}{#1}}
\author{Ziqin Yue}
\affiliation{Department of Physics and Astronomy, Rice University, Houston, Texas 77005, USA}
\alsoaffiliation{Applied Physics Graduate Program, Smalley-Curl Institute, Rice University, Houston, Texas 77005, USA}
\altaffiliation{These authors contributed equally to this work}
\author{Jianwei Huang}
\affiliation{Department of Physics and Astronomy, Rice University, Houston, Texas 77005, USA}
\altaffiliation{These authors contributed equally to this work}
\email{jwhuang@rice.edu}
\author{Ruohan Wang}
\affiliation{Department of Physics and Astronomy, Rice University, Houston, Texas 77005, USA}
\alsoaffiliation{Current address: Department of Physics, Stanford University, Stanford, California 94305, USA}
\altaffiliation{These authors contributed equally to this work}
\author{Jia-Wan Li}
\affiliation{Guangdong Provincial Key Laboratory of Magnetoelectric Physics and Devices, Center for Neutron Science and Technology, School of Physics, Sun Yat-Sen University, Guangzhou 510275, China}
\author{Hongtao Rong}
\affiliation{Institute of Physics, Chinese Academy of Sciences, Beijing 100190, China}
\alsoaffiliation{Current address: Department of Physics, The Pennsylvania State University, University Park, PA 16802, USA}
\author{Yucheng Guo}
\affiliation{Department of Physics and Astronomy, Rice University, Houston, Texas 77005, USA}
\author{Han Wu}
\affiliation{Department of Physics and Astronomy, Rice University, Houston, Texas 77005, USA}
\author{Yichen Zhang}
\affiliation{Department of Physics and Astronomy, Rice University, Houston, Texas 77005, USA}
\author{Junichiro Kono}
\affiliation{Department of Physics and Astronomy, Rice University, Houston, Texas 77005, USA}
\alsoaffiliation{Department of Electrical and Computer Engineering, Rice University, Houston, Texas 77005, USA}
\alsoaffiliation{Department of Materials Science and NanoEngineering, Rice University, Houston, Texas 77005, USA}
\author{Xingjiang Zhou}
\affiliation{Institute of Physics, Chinese Academy of Sciences, Beijing 100190, China}
\author{Yusheng Hou}
\affiliation{Guangdong Provincial Key Laboratory of Magnetoelectric Physics and Devices, Center for Neutron Science and Technology, School of Physics, Sun Yat-Sen University, Guangzhou 510275, China}
\author{Ruqian Wu}
\affiliation{Department of Physics and Astronomy, University of California Irvine, Irvine, California 92697, USA}
\author{Ming Yi}
\email{mingyi@rice.edu}
\affiliation{Department of Physics and Astronomy, Rice University, Houston, Texas 77005, USA}

\title{Topological Surface State Evolution in Bi$_2$Se$_3$ via Surface Etching}

\begin{document}






\begin{abstract}
Topological insulators are materials with an insulating bulk interior while maintaining gapless boundary states against back scattering. \jh{\BiSe~is a prototypical topological insulator with a Dirac-cone surface state around $\Gamma$. Here, we present a controlled methodology to gradually remove Se atoms from the surface Se-Bi-Se-Bi-Se quintuple layers, eventually forming bilayer-Bi on top of the quintuple bulk.} Our method allows us to track the topological surface state and confirm \jh{its robustness} throughout the surface modification. \jh{Importantly, we report a relocation of the topological Dirac cone in both real space and momentum space, as the top surface layer transitions from quintuple Se-Bi-Se-Bi-Se to bilayer-Bi. Additionally, charge transfer among different surface layers is identified.} Our study provides a precise method to manipulate surface configurations, allowing for the fine-tuning of the topological surface states in \BiSe, which represents a significant advancement towards nano-engineering of topological states.
\end{abstract}

\jh{KEY WORDS: Topological insulator, Bi$_2$Se$_3$, ARPES, in-situ etching, surface modification}
\newpage
Topological insulators (TIs) have been a fascinating subject of research in condensed matter physics due to their intriguing physical properties and immense potential for spintronics applications. A canonical topological insulator is characterized by an electrically insulating interior and conducting boundary states, which are guaranteed by the topological character of the bulk electron wavefunctions~\cite{ Kane2005, KaneII2005, Bernevig2006, Moore2007, Konig2007,  FuII2007, Fu2007, Roy2009}. The topological boundary states are protected by time-reversal symmetry (TRS), thus robust against crystal defects that preserve TRS~\cite{Qi2011, Hasan2010}. One prominent characteristic of the topological boundary states is spin-momentum locking, which prohibits backscattering of the boundary conducting electrons from disorders that preserve TRS and holds great promise for potential applications in electronic and spintronic devices~\cite{Moore2010, li2014electrical, jiang2016enhanced}.

As a prototypical three-dimensional TI, the electronic structure of \BiSe~features a large bulk band gap and a Dirac cone topological surface state inside the band gap~\cite{zhang2010crossover, Zhang2009, Xia2009}. Over the past years, tremendous theoretical and experimental efforts have been devoted to understanding the electronic properties of the topological surface states~\cite{Zhang2010, Zhang2009, Xia2009, Park2010, Pan2011, Nechaev2013, Hasan2010}. One major issue that hinders the application of topological surface states in \BiSe~is the instability of its surface termination, potentially due to the high vapor pressure of Se and various surface contamination~\cite{Edmonds2014}. As a consequence, a natural \BiSe~crystal exposed to air is always $n$-type doped with the Fermi level located in the bulk conduction band~\cite{Analytis2010,kim2012surface,zhang2010crossover,Hsieh2009}. In addition, there have been long-standing debates on the surface terminations of \BiSe~\cite{He2013,Diogo2013}. While Se-termination is a natural consequence of the cleavage plane of the van der Waals force-bonded quintuple layers (QLs) Se-Bi-Se-Bi-Se, Bi-termination was reported for \BiSe~stored in ambient conditions~\cite{Hewitt2014, Edmonds2014}. The resulting electronic structures thus can be quite different.

Recently, there have been reports demonstrating that through surface etching with hydrogen ions (H$^+$), bilayer-Bi (Bi$_2$) can form on top of \BiSe~with the removal of Se atoms from the QL~\cite{Shokri2015, Su2017}. The electronic structure of the Bi$_2$/\BiSe~heterostructure fabricated in this top-down method measured with ARPES agrees well with those grown with molecular beam epitaxy~\cite{Miao2013, Eich2014, wang2013creation, sun2019large}. Since free-standing Bi$_2$ thin film is an insulator, charge transfer from Bi$_2$ to \BiSe~was considered to be responsible for the additional bands that appeared around the Fermi level in Bi$_2$/\BiSe~\cite{Govaerts2014}. Strong hybridization of the electronic structures between Bi$_2$ and \BiSe~has been reported while the topological property of the surface states is still preserved~\cite{sun2019large}. However, previous efforts have primarily focused on the well-formed Bi$_2$ on top of \BiSe, without investigating the detailed evolution of the surface configurations and electronic structures in the transition from pristine \BiSe~to Bi$_2$/\BiSe~heterostructure during surface etching. Such studies would provide valuable insights into the manipulation of the topological surface states for various applications.

In this paper, we present a systematic and finely controlled surface etching process that is demonstrated to modify the topological surface states of \BiSe, as characterized with a combination of angle-resolved photoemission spectroscopy (ARPES) and density-functional theory (DFT) calculations. Specifically, we observed a series of gradually evolving electronic band structures as a function of the surface etching time using ARPES. The measured electronic structures correspond to an evolution of surface termination from the pristine \BiSe~to the Bi$_2$/\BiSe~state. From such a systematic electronic structure provided, the nontrivial Dirac crossing can be robustly distinguished from the topologically trivial surface states, supported by the DFT calculations. \jh{Importantly, both momentum space and real space charge redistributions of the nontrivial Dirac cone surface states are revealed, with the discovery of an intriguing intermediate state Bi-Se-Bi-Se/\BiSe. In addition, a systematic surface charge transfer is verified by the corresponding band shift in this process.} Overall, our work offers a comprehensive understanding of the surface termination and the electronic structure evolution during the etching process in \BiSe, and introduces a simple and robust methodology for controlled modification and characterization of the surface chemistry and electronic structure of quantum materials, paving the way for potential future applications.

\begin{figure}
\includegraphics[width=0.5\textwidth]{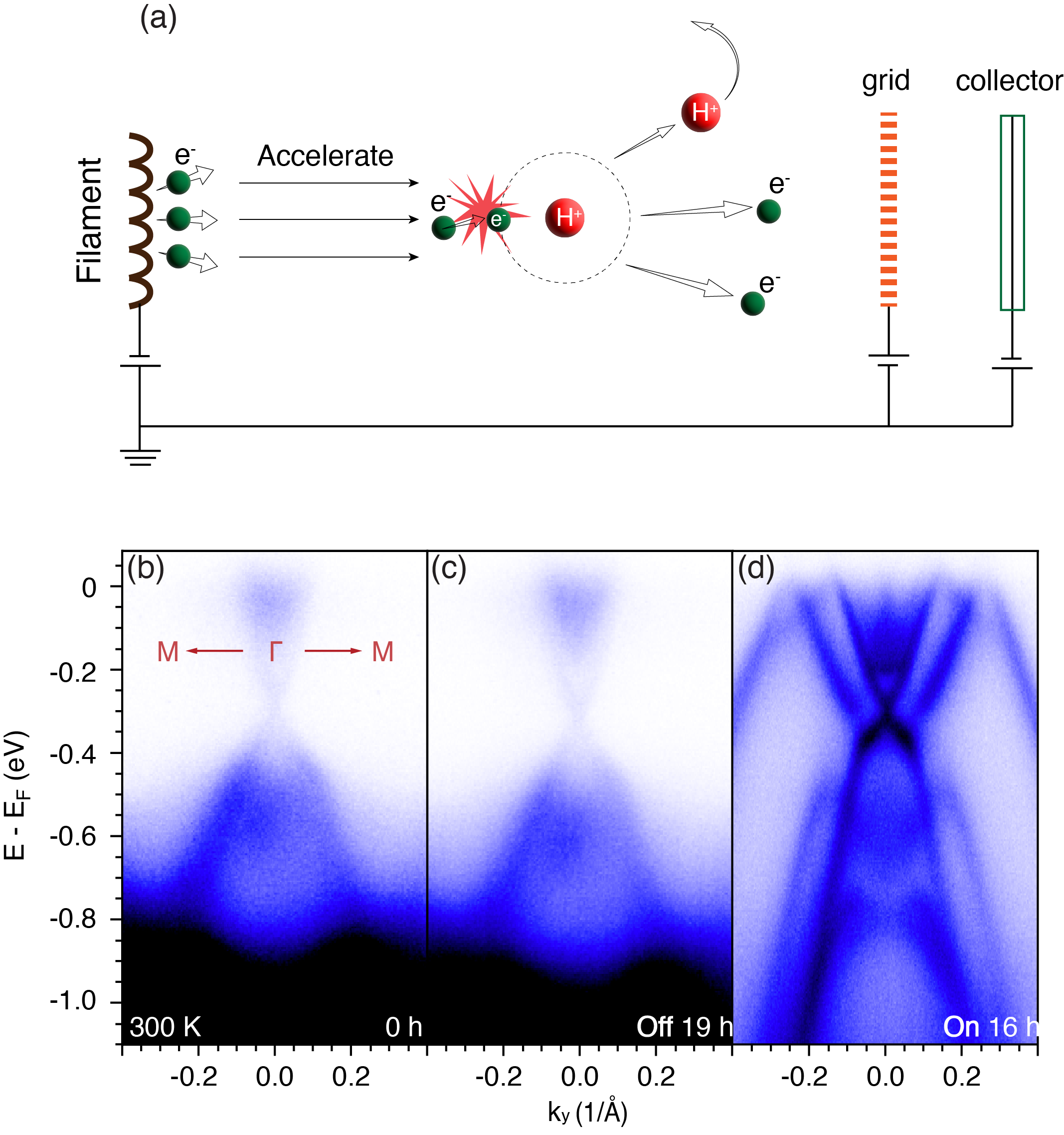}
\caption{\label{fig:onoff} {\bf Surface etching switching on and off using ion gauge.} (a) Schematic showing the generation of H$^+$ out of the residual H$_2$ gas in UHV chamber with the ion gauge on. (b) The band structure of pristine-cleaved \BiSe~measured along M-$\Gamma$-M. (c) The band structure along the same direction after staying in the etching chamber for 19 hours with the ion gauge off. (d) Same as (b) but for 16 hours with the ion gauge on.}
\end{figure}

\begin{figure*}
\includegraphics[width=0.95\textwidth]{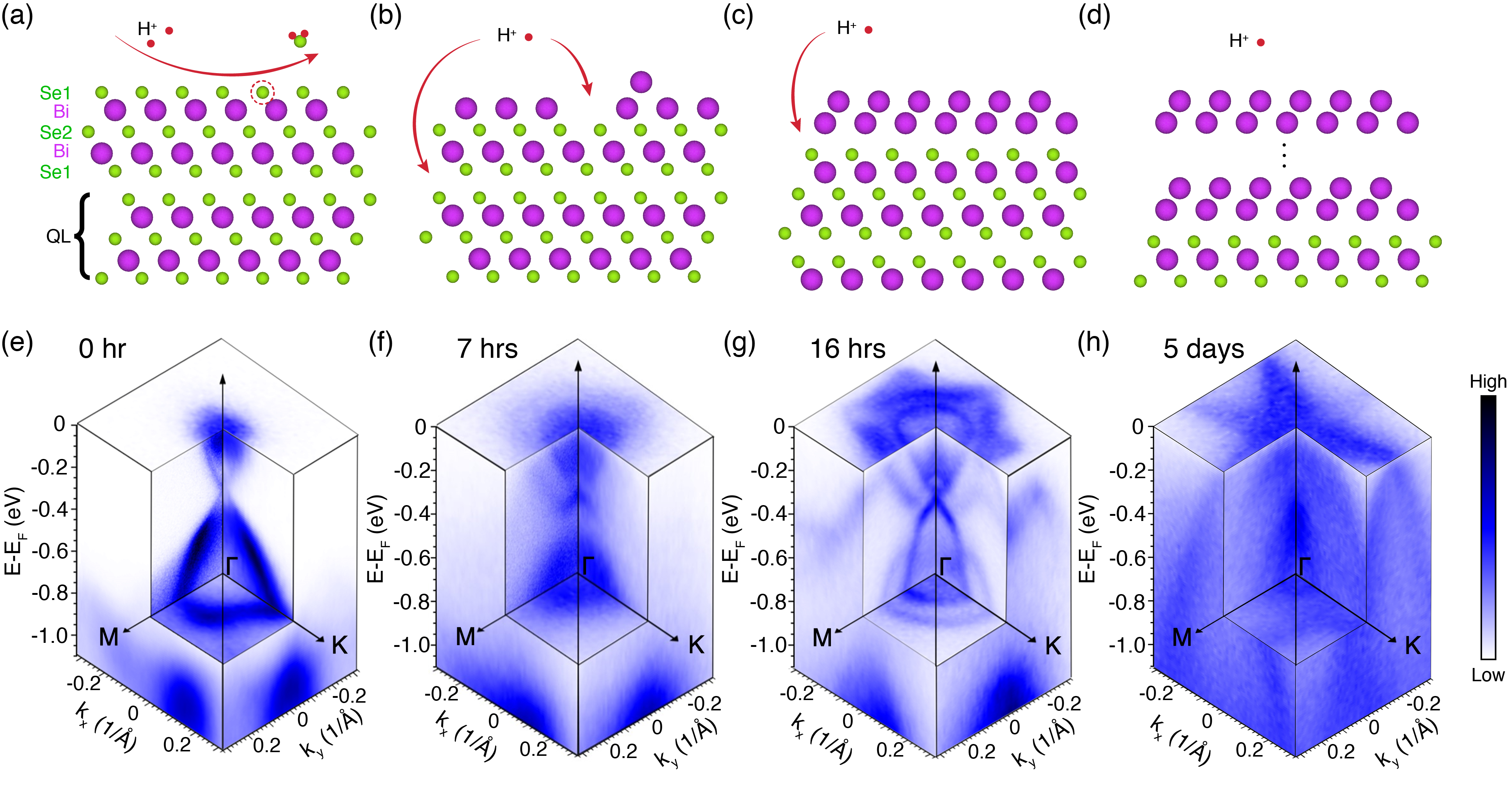}
\caption{\label{fig:schematic} {\bf  Schematic and overview of the evolution of the surface configuration and electronic structure.} (a) The pristine \BiSe~and the H\textsuperscript{+} etching process which initially removes the topmost Se atoms. (b) Complete removal of the topmost Se atoms and the subsequent removal beneath the Bi atoms. (c) Formation of a Bi$_2$ layer on top of \BiSe~QLs. (d) Formation of multiple Bi$_2$ layers on top of \BiSe~through long-time H\textsuperscript{+} etching. (e) Three-dimensional view of the electronic structure of pristine-cleaved \BiSe~measured by ARPES. (f-h) Band structure of surface-etched \BiSe~at different stages. An intermediate etching of 7 hours with removal of topmost Se from the first QL (f); Formation of Bi$_2$ on top of \BiSe~by 16 hours of etching (g); Formation of multiple Bi$_2$ layers after ectching for 5 days (h).
}
\end{figure*}

We have performed a series of surface etching on a freshly cleaved \BiSe~sample in an ultra-high vacuum (UHV) chamber connected to an ARPES measurement chamber. We first introduce our in-situ etching methodology. The mechanism of our etching process is realized through the reaction of the Se atoms in the surface \BiSe~QL with H$^+$ ions generated by the ionization of the residual H$_2$ gas in the chamber activated by the high voltage source of an ion gauge in the UHV chamber (Fig.~\ref{fig:onoff}). \jh{The same effect can be achieved by using a mixture of Ar/H$_2$(3\%) gas in a UHV chamber instead of residual H$_2$ gas (see Fig. S1 for details).} The etching process is controlled by the amount of time that the ion gauge is on and the electronic structure of the \BiSe~sample is measured at the end of each etching step. This methodology is demonstrated to be effective as shown in Fig.~\ref{fig:onoff}. With the ion gauge turned off, leaving the sample in the UHV chamber does not change the electronic structure, but with the ion gauge on, the electronic structure of \BiSe~is dramatically modified as compared to the freshly cleaved state, indicating a modified surface chemistry.

\begin{figure}
\includegraphics[width=0.7\textwidth]{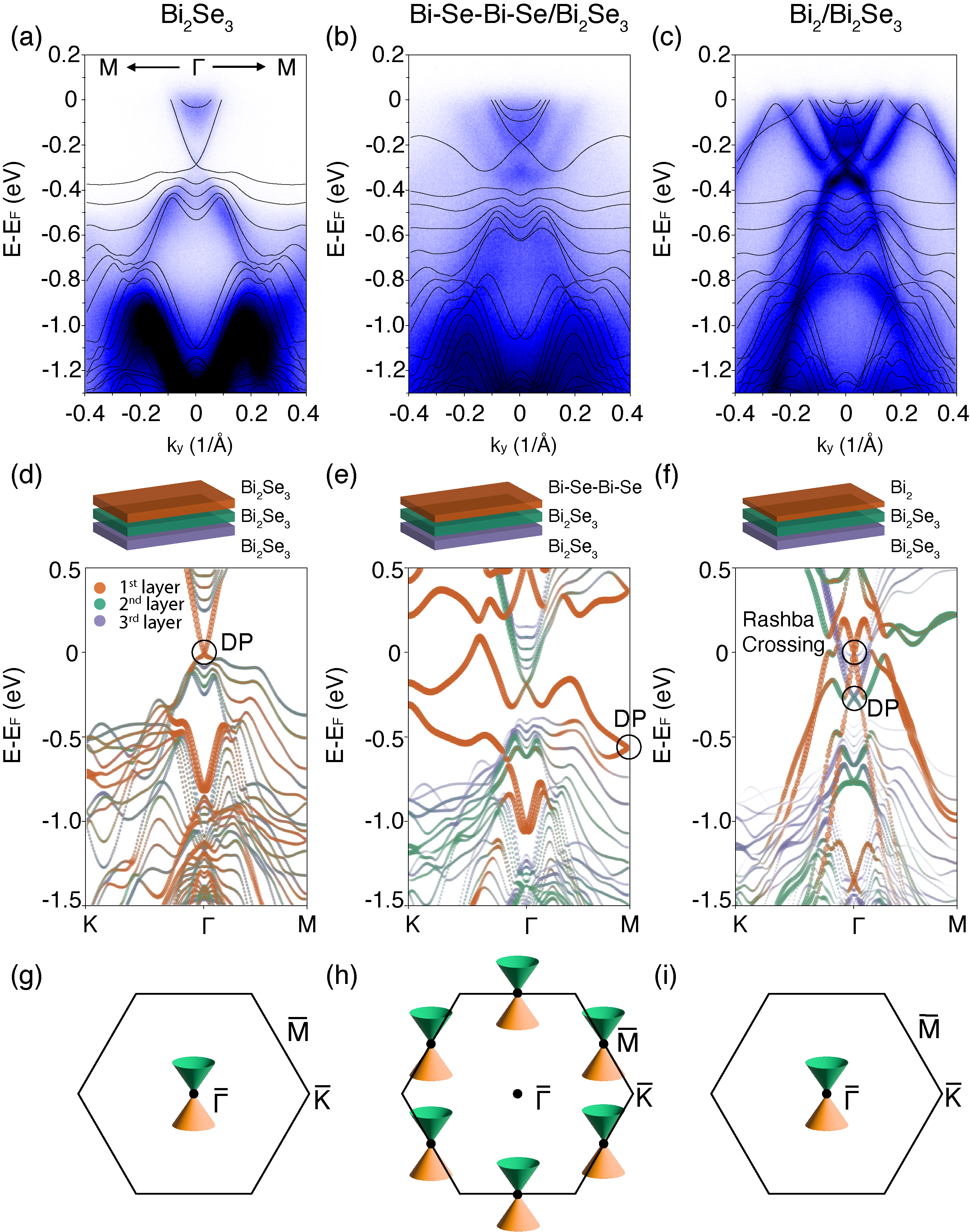}
\caption{\label{fig:calculation} {\bf Evolution of topological Dirac cones of \BiSe~with different surface terminations.} (a-c) Measured ARPES spectral images along M--$\Gamma$--M with the corresponding DFT calculated band structures overlaid at three typical surface etching stages, \BiSe~(a), Bi-Se-Bi-Se/\BiSe~(b), and Bi$_2$/\BiSe~(c). In (a), the calculated band structure was shifted downwards by 0.28 eV to match with the experiments. (d-f) Band structures along K--$\Gamma$--M obtained by DFT calculations of the corresponding surface terminations in (a-c) projected onto the first layer (brown circle), second layer (green circle), and third layer (purple circle). Specifically, the first layer means Bi-Se-Bi-Se layer in (e) while it means Bi$_2$ layer in (f), as shown in the schematic above. (g-i) The momentum positions of the Dirac points corresponding to the different surface terminations in (a-c).
}
\end{figure}

Based on this methodology, we have performed a series of sequential surface etching processes on a freshly cleaved \BiSe~sample and measured the electronic structure immediately after each etching step using ARPES. We first summarize the four states that we have achieved in this process. The surface termination and corresponding measured electronic structure of four selective states are summarized in Fig.~\ref{fig:schematic}, showing the four representative stages achieved after 0, 7, 16 hours, and 5 days of etching, respectively. The freshly cleaved sample in UHV has a \BiSe~QL layer termination as the natural cleavage plane occurs between the QL layers with the van der Waals bonds (Fig.~\ref{fig:schematic}a). The electronic structure measured by ARPES in Fig.~\ref{fig:schematic}e agrees well with previous reports on freshly cleaved \BiSe, showing the well-known Dirac cone surface state inside the bulk band gap~\cite{Xia2009,Park2010,Pan2011}. After a total of 7 hours of cumulative etching, Rashba-like surface states around the Brillouin zone (BZ) center are observed instead of the Dirac cone band dispersion observed in the pristine state (Fig.~\ref{fig:schematic}f). In this case, the top-most Se atoms are almost completely removed, leaving a Se-Bi-Se-Bi surface termination (Fig.~\ref{fig:schematic}b). The Fermi surface and band spectral images are slightly broadened likely due to random distribution of the Se vacancies in the second and third Se layers of the first QL due to etching (see also Fig. S2). The measured Fermi surface and band spectra become sharp again after a cumulative 16 hours of etching (Fig.~\ref{fig:schematic}g). The band structure now becomes more complex with "rose"-like Fermi surfaces, consisting of multiple flower-like and circular Fermi pockets around the BZ center (see also Fig. S3). The electronic structure is nicely reproduced by our theoretical calculations of Bi$_2$/8QL-\BiSe/Bi$_2$ heterostructure (Fig.~\ref{fig:calculation}c), which will be discussed in detail later. The restoration of spectral sharpness indicates the uniformity of the Bi$_2$ layer formation on top of \BiSe. After 5 days of etching, the Fermi surface evolves into a hexagram with a three-fold rotational symmetry in spectral intensity, which corresponds to the formation of multiple Bi$_2$ layers on top of \BiSe~(see also Fig. S4). The spectra become slightly broad again due to the long-time adsorption of gas molecules on the sample surface. The electronic structure of the heterostructures of single and multi-Bi bilayers on \BiSe~have been reported previously as achieved via epitaxial growth of Bi thin film on \BiSe~\cite{Miao2015, tong2020enhanced}. These correspond to the third and fourth stages shown in our Fig.~\ref{fig:schematic}, and hence confirming the effectiveness of our surface etching method on \BiSe.

\begin{figure*}
\includegraphics[width=0.98\textwidth]{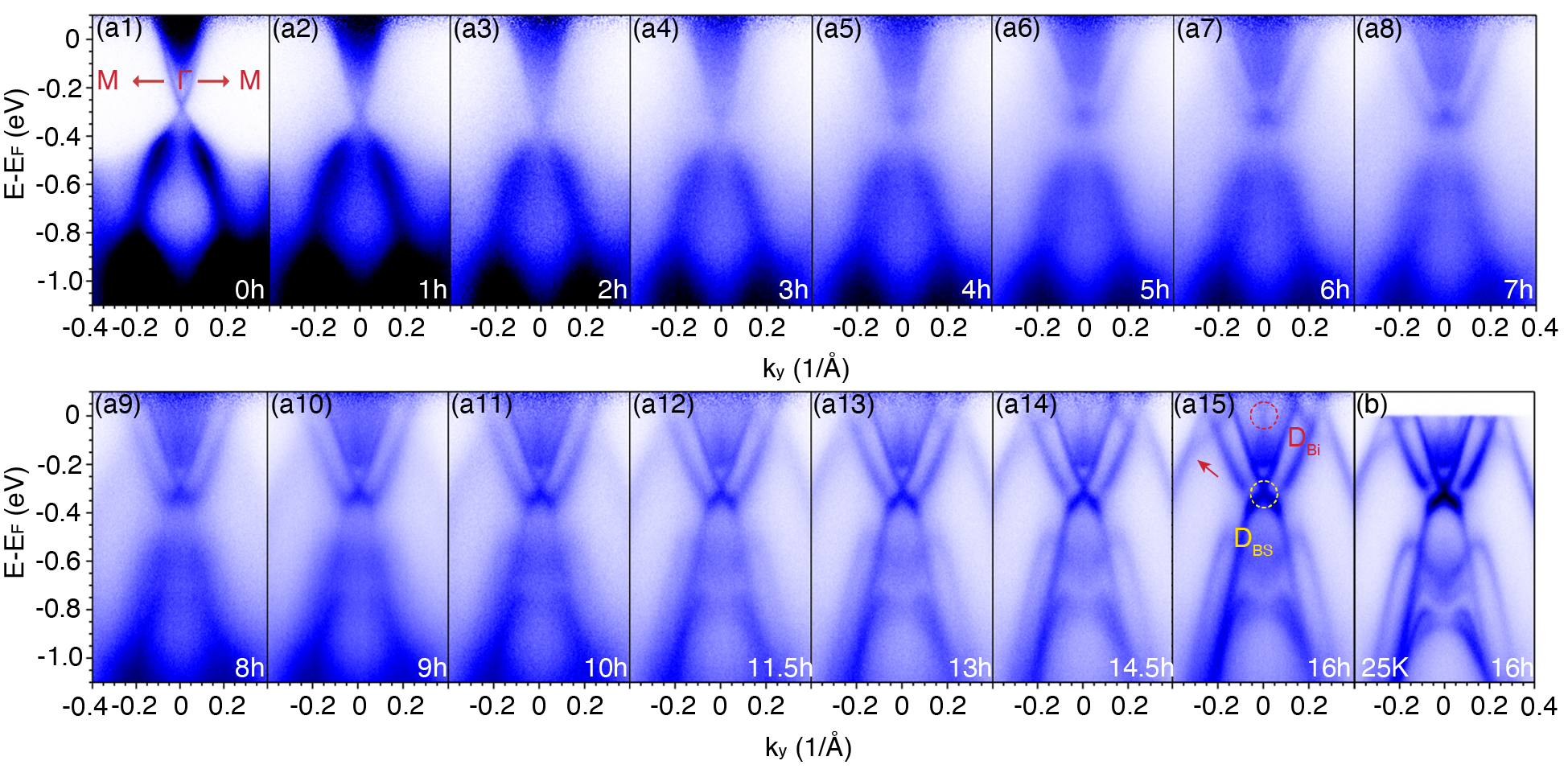}
\caption{\label{fig:evolution} {\bf Detailed ARPES spectra evolution through etching.} (a1-a15) Spectral images of \BiSe~measured by ARPES along the M-$\Gamma$-M direction at 300 K at sequential surface etching steps. The corresponding accumulated surface etching time is indicated at the bottom right of each panel. The yellow dashed circle indicates the Dirac point originating from \BiSe. The red dashed circle and red arrow indicate the corresponding Dirac crossing and bands originating from the top Bi$_2$ layer. The Fermi-Dirac function convolved with the experimental resolution has been divided out from the spectra to reveal the states immediately above the Fermi level due to thermal population. (b) Spectral image along the same direction with 16 hours of surface etching measured at 25 K.
}
\end{figure*}
Having confirmed the surface etching effect on \BiSe, we now address the method we use as a fine-tuning process, which allows for precise control of the topological surface states. A detailed band structure evolution from pristine \BiSe~to a single Bi$_2$ layer on \BiSe~after a total of 16 hours of etching is presented in Fig.~\ref{fig:evolution} (see Fig. S5 for band dispersions in a larger energy range). With the gradual removal of Se during the first 3 hours, the formation of the randomly distributed Se vacancies in the topmost Se layer causes a broadening of the band spectra (Fig.~\ref{fig:evolution}a1-a4). With further etching (4$\sim$8 hours), most of the topmost Se atoms are gone and mesoscopic areas of Bi-Se-Bi-Se regions start to form out of the first \BiSe~QL. The corresponding Rashba-like electronic states appear, replacing the original Dirac cone surface states (Fig.~\ref{fig:evolution}a5-a9). Here, the Rashba-like surface states remain topological, which will be addressed later when combined with DFT calculations. After sufficient etching time (9$\sim$16 hours), more Se atoms from the second and third Se layers of the first \BiSe~QL are removed, and regions of Bi$_2$/\BiSe~start to form. The original Dirac cone surface states emerge again (dashed yellow circle in Fig.~\ref{fig:evolution}a15). Additionally, the formation of the Bi$_2$ layer introduces extra hole-like metallic bands at larger momentum (marked by the red arrow). A second Dirac crossing is also resolved at the Fermi level (dashed red circle in Fig.~\ref{fig:evolution}a15). The measured band structure of the 16-hour-etched \BiSe~is in good agreement with reports for a single Bi$_2$ layer epitaxially grown on the \BiSe~substrate by molecular beam epitaxy method~\cite{Miao2013, Eich2014}. This indicates that a well-defined Bi$_2$ layer has been formed through this time-controlled etching method.
We note the measured spectra have restored spectral sharpness compared with the intermediate Bi-Se-Bi-Se termination state, suggesting a more homogeneous bilayer-Bi that has formed on top of the \BiSe. In addition, Dirac crossing survives during the whole surface etching process, emphasizing the robustness of this time-reversal symmetry protected topological surface state. \jh{Our fine-tuning etching process has enabled a detailed study of surface state evolution in \BiSe~and revealed a new state that has not been reported before.}

To better understand the band structure evolution with the corresponding surface termination development, we now compare the measured ARPES band dispersions with DFT calculations for the corresponding terminations in the sequential stages (Fig.~\ref{fig:calculation}a-c). The DFT calculations were performed for a 6-QL \BiSe, heterostructure with Bi-Se-Bi-Se on a 6-QL \BiSe, as well as heterostructure with Bi$_2$ on a 8-QL \BiSe, to compare with the pristine, 7-hour, and 16-hour etching stages, respectively (Fig.~\ref{fig:calculation}d-f, see also Fig. S6 for more details). To understand the origin of the main features in the calculated band structure, we also show the projection of the calculated band structure unto the top three layers for each structure. All the calculations show good agreement with the corresponding experimental results (Fig.~\ref{fig:calculation}a-c). Specifically, the calculated band dispersion of 6-QL \BiSe~is shifted downward in energy by 0.28 eV to match with the experimental result (Fig.~\ref{fig:calculation}a and d). This can be explained by the tendency of pristine \BiSe~to be electron-doped due to the formation of Se vacancies. A gapless Dirac cone is revealed from both calculation and experiment. For the intermediate etching with the topmost Se layer removed from pristine \BiSe, there are Rashba-like electronic states around the BZ center, which are captured by the corresponding calculations (Fig.~\ref{fig:calculation}b). Usually, a Rashba surface state is considered to be topologically trivial, as it forms a gapped electronic structure around $\Gamma$~\cite{bychkov1984properties,wang2013creation, Govaerts2014}. In this case, the lower branch of the Rashba-like surface state extends to the time-reversal invariant M point, forming a four-fold degenerate Dirac point at M with another band emerging from the valence states (Fig.~\ref{fig:calculation}e). This indicates that the intermediate Bi-Se-Bi-Se/\BiSe~to be still topological but with a Dirac point located at M (Fig. S7). \jh{However, we were not able to observe the Dirac point at M directly using ARPES.} For Bi$_2$ terminated \BiSe, the electronic structure around the Fermi level becomes more complex due to the hybridization of the electronic states from Bi$_2$ and the \BiSe~underneath. Two Dirac cone-like bands are resolved from both experiments and calculations (Fig.~\ref{fig:calculation}c and f). One is located near the Fermi level, as can also be more clearly seen in the Fermi-Dirac function divided spectra in Fig.~\ref{fig:evolution}. This Dirac cone is mainly originating from the Bi$_2$ layer on top. The other Dirac cone is at around -0.3 eV, mainly originating from the first and second \BiSe~QLs. The lower Dirac cone has been recognized to be topological, while the upper one is part of the large Rashba split bands originating from the Bi$_2$ layer, including one of the outside hole-like bands~\cite{wang2013creation}. The electronic structures of the three typical stages reveal that the topological surface state remains robust for the whole etching process, regardless of the surface terminations, which manifests the topological property of the bulk crystal. However, the momentum position of the Dirac point evolves from $\Gamma$ to M and back to $\Gamma$ in this process.

\jh{There are two possible reasons why we were unable to observe the Dirac point at the M point in the Bi-Se-Bi-Se/\BiSe~structure. First, the Dirac point might be strongly suppressed due to our experimental geometry. This hypothesis could be tested by varying the photon energy or light polarization. Second, the Dirac point at M might be gapped by a spontaneous ferromagnetic order induced by Bi$_{Se}$ antisite defects\cite{Nahas2020}. These defects are more likely to form during the etching process. Further investigations are necessary to clarify this issue.}

\begin{figure}
\includegraphics[width=0.7\textwidth]{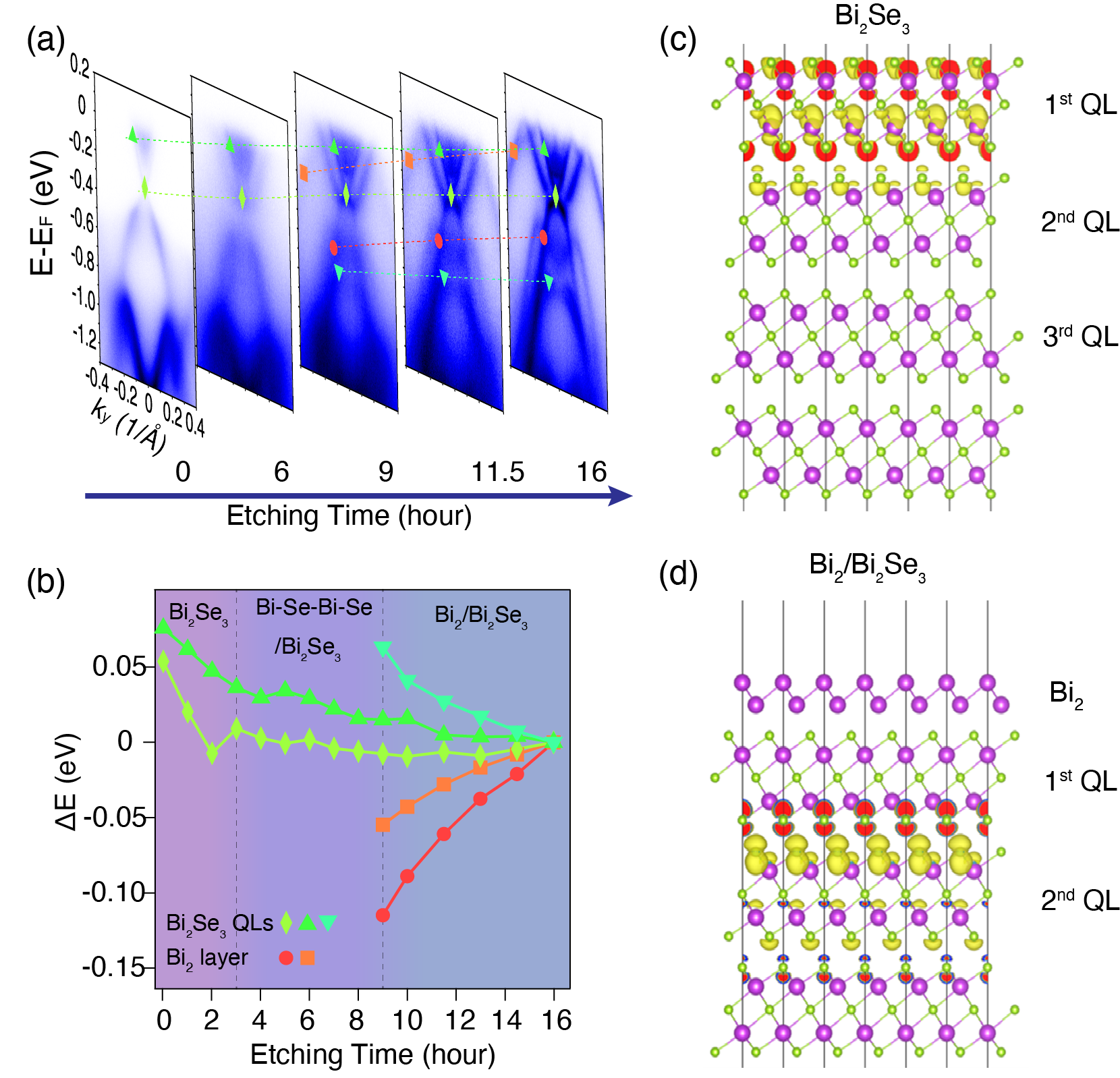}
\caption{\label{fig:EDC} {\bf Real space charge transfer manifested by the shifting of bands originating from different layers.} (a) Selected spectral images measured by ARPES at different surface-etching stages. Different markers indicate different band positions that are being tracked. Dotted curves are drawn as guidelines for tracking the upward and downward shifting of the corresponding bands. (b) Band shift in energy relative to the final 16-hour state as a function of etching time, extracted from energy distribution curve fitting. The green (red) color scheme symbols represent the bands originating from \BiSe~QLs (Bi$_2$). (c) The side view of the pristine \BiSe~lattice and the charge distribution of the Dirac point projected onto different layers obtained by DFT calculations. (d) Same as (c) but of the Bi$_2$/\BiSe~configuration.
}
\end{figure}

We also observe a prominent charge transfer that occurs between the surface layer and the \BiSe~QLs underneath during the surface etching process. This is dictated by the gradual energy shift of the corresponding bands originating from the different surface layers. To demonstrate this, we tracked the energy shift of different bands by fitting the corresponding peak positions from the energy distribution curves (EDCs) as a function of etching time (Fig.~\ref{fig:EDC} and Fig. S8). Distinct charge transfer behaviors at the three different stages are revealed -- electron carriers accumulation in the first \BiSe~QL in the first stage, no significant charge transfer in the second stage and charge transfer between Bi$_2$ and the first \BiSe~QL underneath in the third stage (Fig.~\ref{fig:EDC}b). In the first stage (1$\sim$3 hours), the Se vacancies gradually increase in the first \BiSe~QL, but are largely distributed randomly. The classic Dirac cone dispersion persists, shifting downwards (the green triangles and diamonds in Fig.~\ref{fig:EDC}b) due to the electron doping to the first \BiSe~QL originating from the formation of Se vacancies. As the Se from the surface layer is depleted, and large regions of the Bi-Se-Bi-Se/\BiSe~termination start to form, the system enters the second stage (3$\sim$9 hours). The bands tracked mainly originate from the first \BiSe~QL underneath the surface Bi-Se-Bi-Se layer and do not shift significantly, suggesting minimal to no charge transfer to the first \BiSe~QL layer underneath. During the third stage, large regions of Bi$_2$ start to form on top of the \BiSe~(9$\sim$16 hours). In this stage, new bands emerge that originate from the top Bi$_2$. The  Bi$_2$ bands shift upwards in energy as a function of etching time (red squares and circles in Fig.~\ref{fig:EDC}b), while the bands originating from the first (green reversed triangles in Fig.~\ref{fig:EDC}b) and second (the green triangles and diamonds in Fig.~\ref{fig:EDC}b) \BiSe~QLs beneath Bi$_2$ shift downwards, suggesting an electron charge transfer from the top Bi$_2$ layer to the \BiSe~QLs. 

Another interesting phenomenon is the redistribution of topological surface states in real space from pristine \BiSe~to the formation of Bi$_2$/\BiSe. In Fig.~\ref{fig:calculation}d-f, We note the layer projection of the topological Dirac point evolves from the surface layer to the lower layers underneath. To directly visualize this, we plot the real-space distribution of the electron density associated with the topological Dirac point for the two different surface configurations: pristine \BiSe~and Bi$_2$/\BiSe~(Fig.~\ref{fig:EDC}c,d). From this it is clear that the charge carriers of the nontrivial Dirac point is mainly concentrated in the top termination layer in pristine \BiSe~while they are relocated to the region between the first and second \BiSe~QL beneath the top Bi$_2$ layer in Bi$_2$/\BiSe. Hence during the etching process, the topological electronic states are shifted from the first \BiSe~QL in pristine \BiSe~to the second \BiSe~QL in Bi$_2$/\BiSe. Due to the protection of the 
Bi$_2$ layer on top, the topological electrons tend to be less affected by disorder scattering caused by the absorption of gases on the sample surface.

To summarize, through the combination of ARPES and DFT calculations, we have investigated the electronic structure evolution of \BiSe~under a finely tuned surface etching process, resulting in gradually evolving surface terminations. The etching method was achieved through the chemical reaction of the target material with H$^{+}$ ions, as easily generated by ionizing the residual hydrogen in any UHV chamber equipped with an ion gauge. This simple methodology allows us to gradually remove the Se atoms from the \BiSe~sample surface, and to follow the distinct electronic structures as a function of etching time using ARPES. Starting from the pristine \BiSe, we were able to achieve Bi-Se-Bi-Se/\BiSe~and Bi$_2$/\BiSe~surface terminations. Importantly, the topologically protected surface state of \BiSe~persists with different surface terminations during the whole etching process, proving its robustness against perturbations preserving time-reversal symmetry. However, both the momentum positions and the real space charge distributions of the topological Dirac cones change accordingly in this process. Additionally, real space charge transfer between different surface layers is demonstrated by tracing the shifting of bands originating from different atomic layers. In conclusion, our study provides a simple and finely controlled methodology for manipulating the surface configurations and the corresponding surface states on different quantum material systems via the reaction of Se in the top layer, and could lead to future TI-based spintronics devices.

\section{Author contributions}
Z.Y., J.H. and R. Wang contributed equally to this work.
\section{Author contributions}
J.H. and M.Y. conceived the project. H.R. grew the crystal under the guidance of X.J.Z. J.W.L. performed the calculations under the guidance of Y.H. and R. Wu. J.H., Z.Y., and R. Wang. conducted the ARPES experiments with the help from Y.G., H.W., Y.Z., and M.Y. Z.Y., R. Wang, and J.H. analyzed the data. J.H., Z.Y., R. Wang, and M.Y. wrote the manuscript with input from all authors.

\section{Supporting information}
Experimental details, first-principles calculation methods, additional ARPES data across different etching states, evolution of large energy range ARPES spectra during etching, large energy range calculation results, and etching results using an Ar/H$_2$ mixture.

\begin{acknowledgement}

The ARPES work at Rice University was supported by the Department of Defense, Air Force Office of Scientific Research under Grant No. FA9550-21-1-0343, the Gordon and Betty Moore Foundation's EPiQS Initiative through grant No. GBMF9470 and the Robert A. Welch Foundation Grant No. C-2175. Yusheng Hou was supported by the National Natural Sciences Foundation of China (Grants No. 12104518, 92165204), GBABRF-2022A1515012643, the Open Project of Guangdong Provincial Key Laboratory of Magnetoelectric Physics and Devices (No. 2022B1212010008) and Fundamental Research Funds for the Central Universities, Sun Yat-sen University (No. 24qnpy108).

\end{acknowledgement}



\end{document}